\documentstyle[emulateapj,psfig]{article}

\begin{document}

\submitted{Accepted for ApJ December 1999}

\title{The Local Space Density of Optically--Selected Clusters of Galaxies}

\author{D. A. Bramel, R. C. Nichol \& A. C. Pope}
\affil{Dept. of Physics, Carnegie Mellon University, 5000 Forbes Avenue,
Pittsburgh, PA-15213, USA. }

\begin{abstract}
We present here new results on the space density of rich,
optically--selected, clusters of galaxies at low redshift ($z<0.15$).
These results are based on the application of the matched filter
cluster--finding algorithm (as outlined by Postman et al. 1996 and
Kawasaki et al. 1998) to $1067{\rm deg^2}$ of the Edinburgh/Durham
Southern Galaxy Catalogue (EDSGC). This is the first major application
of this methodology at low redshift and in total, we have detected
$2109$ clusters above a richness cut--off of $R_m\ge50$ (or
$\Lambda_{cl}\geq 10$; Postman et al. 1996). This new catalogue of
clusters is known as the Edinburgh/Durham Cluster Catalogue II (or
EDCCII). We have used extensive Monte Carlo simulations to define the
detection thresholds for our algorithm, to measure the effective area
of the EDCCII and to determine our spurious detection rate.  These
simulations have shown that our detection efficiency is strongly
correlated with the presence of large--scale structure in the EDSGC
data.  We believe this is due to the assumption of a flat, uniform
background in the matched filter algorithm.
Using these simulations, we are able to compute the space density of
clusters in this new survey. We find $83.5^{+193.2}_{-36.9}\times
10^{-6}\,h^{-3}\,{\rm Mpc^{-3}}$ for $100\le R_m<200$
($\Lambda_{cl}\simeq20$) systems, $10.1^{+11.3}_{-4.3}\times
10^{-6}\,h^{-3}\,{\rm Mpc^{-3}}$ for $200\le R_m<400$
($\Lambda_{cl}\simeq40$) systems and $2.3^{+2.5}_{-2.3}\times
10^{-6}\,h^{-3}\,{\rm Mpc^{-3}}$ for $R_m>400$ ($\Lambda_{cl}>80$)
systems.  These three richness bands roughly correspond to Abell
Richness Classes 0, 1 and $\ge2$ respectfully.
These new measurements of the local space density of clusters are in
agreement with those found at higher redshift ($0.2<z_{est}<0.6$) in
the Palomar Distant Cluster Survey (PDCS; Postman et al. 1996 \&
Holden et al. 1999) and therefore, removes one of the major
uncertainties associated with the PDCS as it had previously detected a
factor of $5\pm2$ more clusters at high redshift than expected
compared to the space density of low redshift Abell clusters.  This
discrepancy is now lessened and, at worst, is only a factor of
$4^{+10}_{-4}$. This result illustrates the need to use the same
cluster--finding algorithm at both high and low redshift to avoid such
apparent discrepancies. We also confirm that the space density of
clusters remains nearly constant out to $z\sim0.6$ in agreement with
previous optical and X--ray measurements of the space density of
clusters (Couch et al. 1991; Postman et al. 1996; Ebeling et al. 1997;
Nichol et al. 1999).
Finally, we have compared the EDCCII with the Abell catalogue.  We
detect nearly 60\% of all Abell clusters in the EDCCII area regardless
of their Abell Richness and Distance Classes.  For clusters in common
between the two surveys, we find no strong correlation between the two
richness estimates in agreement with the work of Lumsden et
al. (1992).  In comparison, $\sim90\%$ of the EDCCII systems are new,
although a majority of them have a richness lower than an Abell
Richness Class of 0 and therefore, would be below Abell's original
selection criteria.  However, we do detect 143 new clusters with
$R_m\ge100$ (which corresponds to a Richness Class of greater than, or
equal to, 0) that are not in the Abell catalogue {\it i.e.} 63\% of
the rich EDCCII systems. These numbers lend credence to the idea that
the Abell catalogue may be incomplete, especially at lower richnesses.

\end{abstract}


\section{Introduction}
\label{intro}

Clusters of galaxies play a key role in tracing the distribution and
evolution of mass in the universe (see, for example, Guzzo et
al. 1992; Postman et al. 1992; Nichol et al. 1992, Dalton et al. 1992;
Bahcall \& Soneria 1983; Reichart et al. 1999). Until recently, such
studies have been based on catalogues of clusters constructed from
visual scans of photographic plates {\it e.g.} the Abell catalogue
(Abell 1958; Gunn, Hoessel \& Oke 1986; Abell et al. 1989; Couch et
al. 1991). However, during the past decade, there has been
considerable progress in the construction of automated catalogues of
clusters and groups that possess objective selection criteria. Such
work includes cluster catalogues selected from digitized photographic
material (Dodd \& MacGillivray 1986; Lumsden et al. 1992; Dalton et
al. 1994), from X--ray surveys (Kowalski et al. 1984; Ebeling et
al. 1997; de Grandi et al. 1999; Gioia et al. 1990; Nichol et al. 1997
\& 1999; Rosati et al.  1998; Burke et al. 1997; Jones et al. 1998;
Romer et al. 1999) and large--area optical CCD surveys (Postman et
al. 1996; Lidman \& Peterson 1996; Zaritsky et al. 1997; Olsen et
al. 1999).

\begin{figure*}[tp]
\centerline{\psfig{file=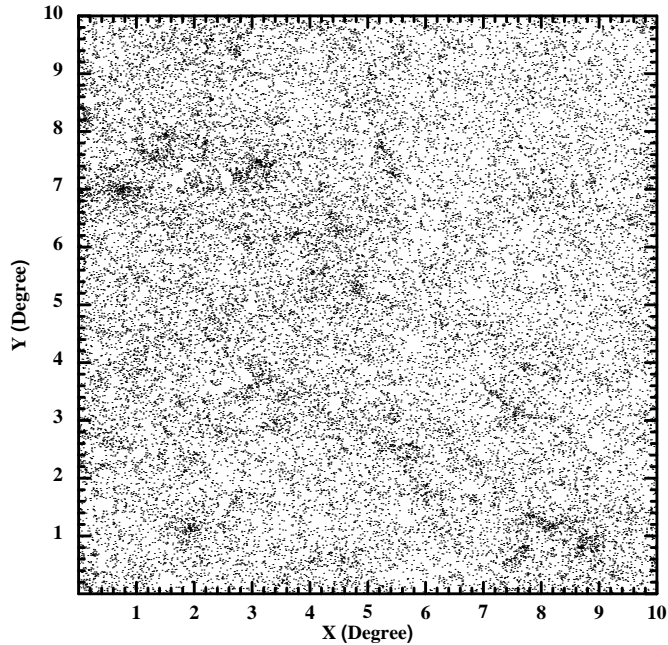,width=3.5in,angle=0}}
\caption{The $10^{\circ}\times10^{\circ}$ test region outlined in
Section \ref{EDSGC} and used for our simulations.  Cluster detections
within $5\times\theta_c$ (at $z=0.05$ this is $0.35^{\circ}$) of the
edges of this data were discarded.
\label{edsgc10x10}
}
\end{figure*}

There has also been great progress in the development of new
cluster--finding algorithms.  The first automated cluster catalogues
used simple variants on the ``peak--finding'' algorithm (Lumsden et
al. 1992) or the percolation method (Dalton et al. 1994).  In recent
years, several new, more sophisticated, algorithms have become
available including the matched filter algorithm -- in several
different flavors (Postman et al. 1996; Kawasaki et al. 1998; Kepner
et al. 1999; Schuecker \& Bohringer 1998) --, the wavelet--filter
(Slezak et al. 1990), the ``photometric redshift'' method (Kodama et
al. 1999), voronoi tessellations (Ramella et al. 1998) and the
``density--morphology'' relationship (Ostrander et al. 1998).  The
level of sophistication of these algorithms has increased in
anticipation of high quality CCD survey data {\it e.g.} the Sloan
Digital Sky Survey (SDSS; Gunn et al. 1998),

The early automated catalogues of optically--selected clusters have
produced two important results.  First, Postman et al. (1996) and
Lumsden et al. (1992) both find evidence for a higher space density of
clusters than that seen in the Abell Catalogue. For example, Postman
et al. (1996) finds that the measured space density of clusters in the
Palomar Distant Cluster Survey (PDCS) is a factor of $5\pm2$ greater
than that implied from the Abell catalogue. Second, the space density
of PDCS clusters remains constant between $z=0.2$ and $z=0.6$, in
agreement with the earlier work of Couch et al. (1991) and has been
confirmed recently by Holden et al. (1999).  If true, these results
can be used to place strong constraints on the underlying galaxy
evolution model ({\it e.g.}  CDM) and measurements of the cosmological
parameters $\sigma_8$ and $\Omega_o$ (see Bahcall, Fan \& Cen 1997;
Reichart et al. 1999; Holden et al. 1999).

To solidify these initial results, larger catalogues of clusters are
required. Moreover, it is becoming increasingly clear that we need to
compare these different cluster catalogues to help verify results and
expand the redshift range over which we can study the cluster
distribution.  To date however, there has been little
cross--comparison between these different cluster
catalogues. Foremost, the relationship between the X--ray and optical
catalogues of clusters remains unclear (see Holden et al. 1997; Briel
\& Henry 1993; Bower et al. 1997).  In the optical domain, different
catalogues have used different cluster finding algorithms thus making
it very difficult to cross--calibrate catalogues and methods and thus
verify results. This is illustrated by the fact that although both
Lumsden et al. (1992) and Postman et al. (1996) find a higher space
density than the Abell catalogue, the PDCS finds 5 times as many
clusters per unit volume, while the Edinburgh--Durham Cluster
Catalogue (EDCC; Lumdsen et al. 1992) only finds twice as many
clusters per unit volume as Abell.  Therefore, it is impossible to
fairly compare the EDCC and the PDCS even though they are both
objective, automated catalogues of clusters.

In this paper, we set out to rectify this problem by running a variant
of the PDCS cluster--finding algorithm on the same galaxy data as used
by Lumsden et al. (1992) in the construction of the EDCC.  The main
aim of this project is to provide a coherent set of cluster data that
spans from $z\sim0.05$ -- the lower redshift limit of the EDSGC -- to
$z\simeq 0.6$ -- the upper completeness limit of the PDCS.  In
addition to using a similar algorithm as Postman et al. (1996), we
have performed a large number of Monte Carlo simulations to assess the
completeness limit, and contamination rate, of this new EDCC cluster
catalogue. This is the first major application of the matched filter
algorithm to low redshift galaxy data, however, it is only the first
of many such surveys presently underway {\it e.g.} the SDSS, DeepRange
(Postman et al. 1998), DPOSS (Gal et al. 1999), COSMOS (Schuecker \&
Boehringer 1998) and the CCD survey of Zaritsky et al. (1997).

In Section \ref{EDSGC}, we discuss the EDSGC catalogue and the matched
filter detection algorithm. In Section \ref{method}, we outline the
methodology used to detect our cluster candidates and discuss in
detail the Monte Carlo simulations we performed to determine our
detection thresholds, the effective area of our new cluster survey and
our spurious detection rate.  Readers interested in just the results
of this survey may wish to concentrate on Section \ref{results} of
this paper which presents our space density results.  In Section 5, we
discuss these results in comparison with the PDCS and Abell cluster
catalogues.  Throughout this letter, we use $H_o=100\,h\,{\rm km
s^{-1} Mpc^{-1}}$ and $q_o=0.5$ unless otherwise stated.

\begin{table*}[pt]
\begin{center}
\begin{tabular}{c|ccccc}
         & \multicolumn{5}{c}{${\rm Redshift}\,\, (z_{est})$}\\
$R_m$	& 0.05	& 0.07	& 0.09	& 0.12	& 0.15 	\\
\hline\hline
50	& 270	& 210	& 190	& 220	& 320 	\\ 
100	& 310	& 270	& 230	& 250	& 320 	\\ 
200	& 410	& 340	& 340	& 350	& 410 	\\ 
400	& 600	& 480	& 460	& 480	& 540 	\\ 
\end{tabular}
\caption{\centerline{Richness threshold values (RT) determined for various
redshifts and cluster richnesses}
\label{richthreshtable}
}
\end{center}
\end{table*}

\section{The Edinburgh--Durham Southern Galaxy Catalogue}
\label{EDSGC}

The Edinburgh--Durham Southern Galaxy Catalogue (EDSGC) has been
discussed in detail in Heydon--Dumbleton et al. (1989), Lumsden et
al. (1992), Collins et al. (1992), and Collins, Nichol, and Lumsden
(2000). However, for consistency, we include here a brief discuss of
the salient points of this catalogue.

The whole EDSGC comprises of ~1.5 million galaxies brighter than
$b_j=21.5$ covering an area of $\sim 1100$ squared degrees centered on
the South Galactic Pole, spanning 90 degrees in Right Ascension and 20
degrees in declination. The catalogue is $95\%$ complete to $b_j=20$
and has $\leq 10\%$ stellar contamination. The catalogue was
constructed from COSMOS (a microdensitometer) scans of UK Schimdt
IIIa--J photographic survey plates and as photometrically calibrated
using 30 CCD sequences positioned in a ``checkerboard fashion''. From
the EDSGC, Lumsden et al. (1992) detected $733$ galaxy overdensities
using a simple ``peak--finding'' algorithm. Collins et al. (1995)
presents 777 redshifts measurements within 94 clusters and this
redshift sample has been used to study the large--scale distribution
of nearby clusters (see Nichol et al. 1992, Guzzo et al. 1992 \&
Martin et al. 1995) as well as the cluster luminosity function
(Lumsden et al. 1997).

\begin{table*}[pt]
\begin{center}
\begin{tabular}{c|ccccc}
         & \multicolumn{5}{c}{${\rm Redshift}\,\, (z_{est})$}\\
$R_m$	& 0.05	& 0.07	& 0.09	& 0.12	& 0.15 	 \\
\hline\hline
50	& -230	& -130	& -90	& -65	& -60 	\\ 
100	& -230	& -150	& -90	& -65	& -65 	\\ 
200	& -250	& -160	& -110	& -80	& -65 	\\ 
400	& -360	& -200	& -140	& -100	& -70 	\\
\end{tabular}
\caption{\centerline{Log likelihood threshold values (LT) determined for various
redshifts and cluster richnesses}
\label{likethreshtable}
}
\end{center}
\end{table*}

For the analysis discussion in Section \ref{method}, we restrict
ourselves to a small $10^{\circ}\times 10^{\circ}$ random subregion of
the EDSGC centered at $00^{hrs}$ $30^{mins}$ in Right Ascension and
$-33^{\circ}$ in Declination. We also restricted the magnitude range
to $15<b_j<20.5$ to remain as complete as possible (see Collins,
Nichol \& Lumsden 2000).  These cuts resulted in a total of 41171
galaxies which is significantly smaller than the whole EDSGC.  This
test data is shown in Figure \ref{edsgc10x10}.  For the space density
results presented in Section \ref{results}, we used to all galaxies in
the magnitude range $15<b_j<20.5$ and in a coordinate range of
$22^{\rm hrs}<\alpha<3.3^{\rm hrs}$ and
$-42^{\circ}<\delta<-23^{\circ}$ ($1067{\rm deg^2}$) which gave us
627260 galaxies in total.

We note here that the EDSGC has previously been used to construct an
objective catalogue of clusters of galaxies (see Lumsden et al. 1992).
However, in this prior analysis, only a simplistic ``peak--finding''
algorithm was used to find candidate systems for redshift follow--up.
Given the recent advances in cluster--finding algorithms, we decided
it was prudent to repeat the analysis which we discuss herein. We
stress however that this does not undermine the scientific integrity
of the original EDCC catalogue (Lumsden et al. 1992) and results
derived from it (Nichol et al. 1992; Collins et al. 1995; Martin et
al. 1995). In this present work, we simply wish to analyze the low
redshift cluster population using the same techniques as presently
used at high redshift ({\it i.e.}  PDCS). For the sake of consistency,
we call this new cluster catalogue the Edinburgh/Durham Cluster
Catalogue II (EDCCII).

\section{Methodology}
\label{method}

\subsection{The Matched Filter}
\label{matched}

In recent years, there has been considerable progress in the
development of new, automated cluster--finding algorithms (see Lumsden
et al. 1992; Dalton et al. 1994; Postman et al. 1996; Kawasaki et
al. 1998; Kepner et al. 1999; Schuecker \& Bohringer 1998; Slezak et
al. 1990; Kodama et al. 1999; Ramella et al. 1998; Ostrander et
al. 1998). In this paper, we focus our attention on the matched filter
algorithm since we are interested in directly comparing our results to
those of Postman et al. (1996).

We have based our matched filter algorithm on the procedure outlined
by Kawasaki et al. (1998) which compares the galaxy distribution
around any point on the sky to a cluster model plus a background (see
Eqn. 1 in Kawasaki et al. 1998).  The parameters of this cluster model
are given in Eqns. 2, 3 \& 4 of Kawasaki et al. (1998) and are the
cluster surface density profile, the intrinsic richness of the cluster
($N$ in Eqn 1. of Kawasaki et al. 1998), the cluster and field
luminosity functions and the surface density of background galaxies
($\sigma_f$).  For the analysis discussed herein, we have modeled our
clusters as a spherically--symmetrical, isothermal surface density
profile ({\it i.e.} a King profile with $r_{core}=170$ kpc and
$\beta=\frac{2}{3}$) combined with a Schechter luminosity function
(with $M^{\ast}_{b_j}=-20.12$ and $\alpha=-1.25$; see Lumsden et
al. 1997).  The background galaxy distribution is modeled as a flat
surface density of galaxies of $\sigma_f=583775$ galaxies per
steradian which is the measured average surface density of galaxies in
the EDSGC in the magnitude range $15<b_j<20.5$. For the field
luminosity function ($\theta_f(m)$ in Eqn. 3 of Kawasaki et al. 1998),
we used a Schtecter function with $M^{\ast}_{b_j}=-19.5$ and an
$\alpha=-1.1$ (see Loveday et al. 1992).

The physical model for the cluster plus background is converted to an
observational model using the standard cosmological redshift--distance
relationships (and k--corrections) and therefore, the model is only a
function of the intrinsic richness of the cluster and its redshift.
This observed model is convolved with the data and a likelihood
assigned for each point on the sky which is proportional to the
quality of fit of the cluster model to the observed galaxy
distribution given Possion statistics (see Eqns 6 \& 7 of Kawasaki et
al. 1998).  One can then maximized the likelihood by varying the two
parameters of the model {\it i.e.} richness and redshift.
Computationally, this was achieved by overlaying the EDSGC with a grid
(pixel scale of $\theta_c/3$), and comparing each pixel, in this grid,
with the matched filter model as a function of cluster redshift and
richness ($R_m$).  For the results presented herein, we varied the
matched filter redshift ($z_{est}$) from $0.05$ to $0.15$ ($\delta
z_{est}=0.0025$) and used richness estimates of $R_m=50$, 100, 200 and
400. This resulted in an array of likelihood and richness maps (see
Kawasaki et al. 1998) from which we must select our cluster
candidates.

\subsection{Monte Carlo Simulations}

In this section, we outline our Monte Carlo simulations which were
used to determine the cluster detection thresholds as well as to
estimate our spurious cluster detection rate and overall cluster
detection efficiency.

\subsubsection{Model of a Cluster}
\label{model}
For our Monte Carlo simulations, we must first create an artificial
cluster.  We used a spherically--symmetrical, isothermal surface
density profile with a cluster core radius ($r_c$) of $170$ kpc, a
cutoff radius of $5 r_c$ and a Schtecter luminosity function with
$\alpha=-1.25$ and $M^{\ast}_{B}=-20.12$ (we restricted ourselves to
an absolute magnitude range of $M^{\ast}_{b_j}\pm 5$).  This
artificial cluster was then redshifted appropriately using the
standard cosmological relations and

\begin{equation}
m = M + (42.384 + 5\log z) + (4.14 z - 0.44 z^2),
\end{equation}

\noindent to convert absolute magnitude to apparent $b_j$ magnitude
(see Lumsden et al. 1997).  To match the magnitude range covered by
the EDSGC survey, we then removed all galaxies outside the magnitude
range of $15<b_j<20.5$.  Four different intrinsic richnesses of
artificial cluster were used in our Monte Carlo simulations;
$R_m=50,\, 100,\, 200,\, \&\, 400$.  Unless otherwise stated, each
artificial cluster was unique, with the galaxies distributed at random
according to the angular and luminosity distributions given above.

We note here that no attempt was made to simulate the
density--morphology relationship or to allocate a cD--type galaxy at
the cluster core. Moreover, we did not change the shape or parameters
of our artificial clusters during our simulations.  Such simulations
would have been computational intensive but are clearly needed in the
future.

more would have resulted in merging of the artificial clusters, while
adding fewer clusters would have made the simulations laborious.  We
then varied the thresholds, both in richness (RT) and likelihood (LT),
and computed for each combination the number of artificial clusters
detected as well as the total number of clusters detected above these
thresholds.

It was then necessary to weight these detections to determine the
optimal thresholds.  Clearly, we wish to rule out obvious cases {\it
i.e.} detecting 1 artificial cluster while detecting hundreds of other
systems within the EDSGC (as most of these detections will be either
lower richness clusters or spurious detections).  Unfortunately, most
cases we encountered were more subtle than this {\it i.e.} detecting
10 artificial clusters within a total of 25 detections, or detecting
15 artificial clusters within a total 40 detections.  To help
differentiate between these intermediate cases, we employed an
analytical method which we outline below.

We defined the number of detections of our artificial clusters to be
$x$ and the number of total detections to be $y$.  Both $x$ and $y$
are functions of RT and LT and increase with decreasing thresholds. We
therefore defined a function (T($x,y$)) which satisfied the following
boundary conditions; T($0,y$)$ = 0$, $\lim_{y \rightarrow \infty}$
T($x,y$) $= 0$ and the maximum of T($x,y$) is at T($20,20$).  The
functional form of T($x,y$) is therefore important since it is now our
weighting scheme.  We estimated T($x,y$) by running our
cluster--finding algorithm over many thousands of realizations of our
simulations and varying RT and LT to create a Monte Carlo estimate for
T$(x,y)$.  From these data, we empirically found that
T($x,y$)$=\frac{x^3}{y^{1.3}}$ was the best functional form for this
data and adopted it as our weighting scheme.  Using this relationship,
we then determined the optimal RT and LT thresholds as a function of
redshift and intrinsic richness.  Our thresholds are given in Tables
\ref{richthreshtable} and \ref{likethreshtable} (we used linear
interpolation between these values if necessary).

\begin{figure*}[pt]
\centerline{\psfig{file=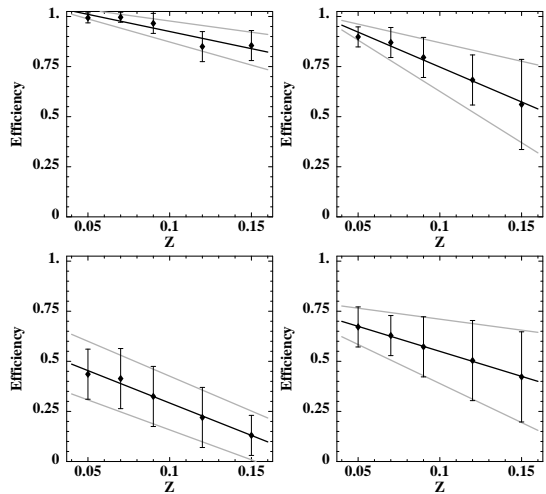,width=3.in,angle=0.}}
\caption{The averaged measured detection efficiency for clusters of
various richness as a function of redshift.  Clockwise from the upper
left--hand panel, the graphs represent clusters of richnesses $R_m=$
400, 200, 100, and 50.
\label{efficiencies}
}
\end{figure*}

\subsubsection{Locating Cluster Candidates}

In this subsection, we review our procedure for identifying a unique
cluster candidate above the thresholds outlined above.  We first
designate all pixels in our likelihood and richness maps that satisfy
our thresholds, $\ge$RT and $\ge$LT, as ``active'' pixels.  Obviously,
a single cluster candidate will create multiple active pixels so we
must group these pixels together into single detection.  This is
achieved by searching for peaks in the distribution of active pixels
{\it i.e.}  we look for a active pixel whose height is greater than
any other active pixel within a radius of $2 {\theta_c}$ of the
peak. We then group together all active points in this area to create
a unique cluster candidate.  If a single active pixel is a peak by
default -- i.e. there are no other active pixels within $2 \theta_c$
of it -- we disregard this peak and do not include it in our final
analysis.  Such isolated peaks are unlikely to be caused by real
clusters.  Finally, we compute the weighted mean of all active points
grouped together as a single cluster to determine the most likely
cluster candidate centroid.

\subsubsection{Determining Detection Efficiencies}
\label{efficiency}

In addition to determining our detection thresholds, our Monte Carlo
simulations were used to estimate our detection efficiency and the
effective area of our cluster search. This is an important aspect of
any cosmological survey as it allows us to determine the volume
sampled by the EDCCII and thus measure the space density of clusters
from the catalogue.  Previous applications of automated optical
cluster--finding algorithms have based their efficiency measurements
on their ability to detect clusters within artificial galaxy data. For
example, Postman et al. (1999) and Kepner et al. (1999) created
artificial galaxy catalogues with the same statistical properties as
real galaxy catalogues {\it e.g.}  they matched the surface density of
galaxies and/or the large--scale clustering properties of the
galaxies. They then added artificial clusters to such simulated galaxy
data.

We however found this method inadequate since such simulated galaxy
catalogues cannot fully reproduce the hierarchical structure of
galaxies in the universe.  Specifically, the effects of superpositions
of structures, variations in the field counts, or large--scale
structures are hard to include in these simulations.  Therefore, for
each combination of RT and LT (see Tables \ref{richthreshtable} and
\ref{likethreshtable}), we added a total 6000 artificial clusters (20
at a time) at random in the EDSGC and computed our success rate in
detecting these artificial clusters above our thresholds (an
artificial cluster was considered detected if a candidate cluster was
found by our algorithm within $2 \theta_c$ of the original coordinate
of the artificial cluster; $\theta_c$ was evaluated at the redshift of
the artificial cluster).

In Figure \ref{efficiencies}, we present our average detection
efficiencies as a function of input redshift and richness (using the
$10^{\circ}\times10^{\circ}$ test area discussed in Section
\ref{EDSGC} and in Figure \ref{edsgc10x10}). The error bars are the
standard deviation observed between the different trials of 20
clusters added to the EDSGC data at any one time.

\begin{figure*}[pt]
\centerline{\psfig{file=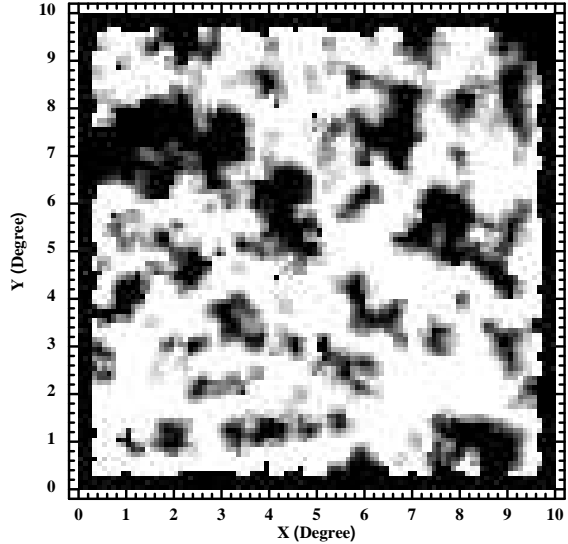,width=3.in,angle=0.}}
\caption{Our detection efficiency as a function of position in the
EDSGC data shown in Figure~\ref{edsgc10x10}.  Light areas indicate a
detection efficiency of $100\%$ for our artificial clusters, while
dark areas indicate a 0\% detection efficiency.  Efficiencies near the
edge of the area were not measured.  This plot is for $z_{est}=0.05$
and $R_m=100$ artificial clusters.
\label{LSSinterfere}
}
\end{figure*}

In Figure \ref{LSSinterfere}, we show an example of the angular
dependence of our detection efficiency. It is interesting to note that
our efficiency is strongly correlated with the large--scale structure
(LSS) in the Universe.  This effect is most prominent for lower
richness clusters, while for the richer systems ($R_m\ge 200$), it is
insignificant (except at the highest redshifts probed by our
simulations where the effect resulted in a loss of $\simeq15\%$ of
artificial clusters).  In addition to this interference, our
efficiency in detecting low richness, high redshift clusters was
hindered by the magnitude limit of the EDSGC data since the number of
potentially visible galaxies in these artificial clusters decreases
below the noise in the background.

This problem of LSS interference appears to effect both over-- and
under--dense regions of the EDSGC data (when compared to the mean
surface density of galaxies).  For example, the most striking example
of this LSS interference is seen at coordinates 2.0, 7.5 in Figure
\ref{edsgc10x10}, where a group of clusters -- Abell 2730, 2721, 2749,
2755, and 12S -- has reduce our detection efficiency to almost zero
(see Figure \ref{LSSinterfere} ).  In contrast, we also observe in
Figure \ref{LSSinterfere} a low detection efficiency near coordinates
8.0, 6.0 which coincides with a under--dense region in Figure
\ref{edsgc10x10}.  We believe this effect is caused by our assumption
of a flat, uniform galaxy background as used in the matched filter
algorithm (see Section \ref{matched}). In both over-- and under--dense
regions, our assumption of a flat background with the mean surface
density of the EDSGC is poor and therefore, we suppress the overall
likelihood of the cluster detection {\it i.e.} the model of the
background around the cluster is not a flat surface density of
$\sigma_f=583775$ galaxies per steradian.  This effect is further
exasperated by systematic plate--to--plate uncertainties in the
magnitude zero-point of the EDSGC photographic plates (Nichol \&
Collins 1993).

\begin{table*}[pt]
\begin{center}
\begin{tabular}{c|ccccc}
         & \multicolumn{5}{c}{${\rm Redshift}\,\, (z_{est})$} (${\rm deg^2}$) \\
$R_m$    & 0.05 & 0.07 & 0.09 & 0.12 & 0.15 \\ \hline\hline
50       & 580  & 516  & 415& 266  & 196 \\
100      & 742  & 687  & 612& 553  & 445 \\
200      & 962  & 928  & 855& 729  & 594 \\
400      & 1067  & 1063  & 1034& 966  & 908 \\ 
\end{tabular}
\caption{The effective area (${\rm deg^2}$) of the EDCCII as a
function of input cluster redshift and richness. These numbers were
extrapolated from the efficiencies we derived for the smaller test
area of the EDSGC.
\label{area}
}
\end{center}
\end{table*}

We present the effective area of the whole EDCCII in Table \ref{area}
based on our simulation results.  These data were obtained by summing
over the larger EDSGC survey area, as defined in Section \ref{EDSGC},
but weighted by our success rate in detecting artificial clusters as
computed from the smaller test area.  The data given in Table
\ref{area} illustrate the power of this new EDCCII catalogue since we
now know the selection function of an optically--selected cluster
catalogue to the same accuracy as a X--ray--selected cluster survey
({\it e.g.}  Nichol et al. 1999). This effective area will be used
below when calculating the space density of clusters (see Section
\ref{results})

\subsubsection{Determining Spurious Detection Rate}
\label{spurious}

The final use of our simulations was to determine the likely spurious
detection rate. Again, we have tried to use the real galaxy data as
much as possible in this analysis so as to mimic the real
uncertainties in the EDSGC catalogue.  This is different from previous
attempts to estimate the spurious detection rate which have relied on
simulated galaxy catalogues (see Postman et al. 1996).

To achieve this goal therefore, we perturbed each galaxy in the EDSGC
in a random distance (between $2 \theta_c$ and $5 \theta_c$ with a
flat distribution) in a random direction from its original position.
This procedure effectively smoothes the galaxy catalogue on these
particular scales removing all small--scale structure in the catalogue
while retaining the large--scale features within the catalogue.  We
then applied our matched filter algorithm to these perturbed galaxy
catalogues and calculated the number of clusters that would satisfy
our selection criteria.  This was performed many thousands of times to
determine the standard deviation in our spurious detection rate.

\begin{figure*}[pt]
\centerline{\psfig{file=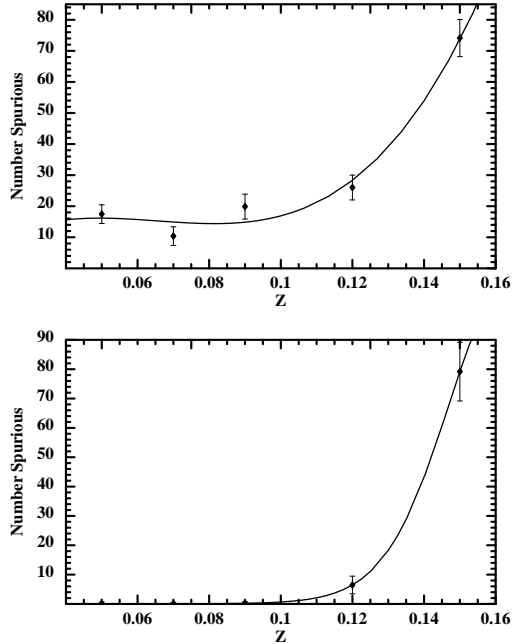,width=3.in,angle=0.}}
\caption{Number of spurious detections as a function of redshift for
$R_m=50$ (top) and $R_m=100$ (bottom) clusters in our 10x10 EDSGC test
area.  Number of spurious detections for higher richness clusters were
negligible {\em i.e.} less than one spurious cluster in the
$10^{\circ}$x$10^{\circ}$ area.
\label{spuriousplot}
}
\end{figure*}

Our spurious detection rates are shown in Figure \ref{spuriousplot}.
The number of spurious detections was significant only for high
redshift $R_m=50$ and $R_m=100$ clusters.  Based on these simulations
therefore, we restrict ourselves to $z\le0.12$ for $R_m\le100$ and
$z\le0.15$ for $R_m>100$ to ensure that the spurious detection rate
remains insignificant. For example, for $R_m\ge200$ systems, $<1\%$ of
detected clusters are likely to be spurious below $z=0.15$.  We also
exclude all clusters at $z<0.05$ as our thresholds are not accurately
calibrated below this redshift.  In Section \ref{results} therefore,
we restrict ourselves to $R_m\ge100$ systems in the redshift range
$0.05<z<0.15$ and make no correction for spurious detections within
this richness and redshift range.

We note here that the results of these simulations are in good
agreement with empirical determinations of the completeness of the
EDSGC and EDCC catalogues {\it i.e.}  based on 777 galaxy redshift
measurements, Nichol (1993) showed that the EDCC was complete out to
$z\simeq0.13$ and only contained $<15\%$ contamination from spurious
clusters.  At low redshift, Lumsden et al. (1992) also had difficulty
detecting $z<0.03$ clusters in the original EDCCI catalogue because of
the large angular size subtended by such clusters.

In addition to randomizing the positions of the EDSGC, we also
performed simulations which randomly shuffled the magnitudes of the
galaxies throughout the EDSGC data.  This resulted in galaxy
catalogues with the same statistical properties -- {\it i.e.} same
angular clustering and number--magnitude relationship -- but removed
all correlations with magnitude.  Again, these randomized catalogues
were analyzed with our matched filter algorithm and resulted in a very
similar result as presented in Figure \ref{spuriousplot} {\it i.e.}
the spurious detection rate was insignificant for lower redshift,
higher richness systems.  We note however, that on average we detected
fewer rich systems than with the real EDSGC data. In other words,
magnitude correlations only appear to aid in the detection of rich
clusters in the data.

\subsubsection{Merging of Catalogues}

Once we have the cluster detections, as a function of redshift and
richness, we must then remove duplicate cluster detections to produce
a final catalogue of unique cluster candidates. This was achieved by
grouping together all systems whose measured centroids were within
$2\,\times\,\theta_c$ of each other.  We began this process by
grouping together candidates as a function of their redshift, followed
by their richness.  If duplicates were found, we simply averaged their
richness and redshift estimates to obtain a final estimate of the
candidate clusters' richness and redshift.

\section{The Space Density of EDCCII Clusters}
\label{results}

\begin{table*}[pt]
\begin{center}
\begin{tabular}{r|cccc}
 &\multicolumn{4}{c}{$N(\geq R_m (\Lambda_{cl}),\leq z_{est})\,\,\,({\rm deg^2})$ } \\
$R_m$ ($\Lambda_{cl}$)	&$0.12$ or $0.15$	&$0.3$	& $0.4$ &$0.6$	\\
\hline\hline
$R_m\ge100$ ($>20.3$)	&$0.41^{+0.72}_{-0.23}$ ($z_{est}\le 0.12$)  &$3.1$& $4.9$ &$7.6$	\\
$R_m\ge200$ ($>40.6$)	&$0.074^{+0.075}_{-0.029}$($z_{est}\le 0.15$) &$1.4$& $3.2$&$5.9$	\\
$R_m\ge400$ ($>81.2$)	&$0.015\pm 0.015$ ($z_{est}\le0.15$)	     &$0.015$ &$0.4$&$1.4$\\
\end{tabular}
\caption{Comparison of cumulative surface densities of the EDCCII and
the PDCS.  The cumulative surface densities of the PDCS clusters were
calculated using the data in Table 4 of Postman et al. (1996) using
the EDCCII as a zero point.  We excluded all PDCS clusters with
$\sigma<3$ and a radius of $>200$ since such clusters have a higher
probability of being spurious (see Postman et al. 1996).  We assumed
$5.1\,{\rm deg^2}$ for the surface area of the PDCS survey.  Our PDCS
surface densities agree with those presented in Figure 21 of Postman
et al. (1996).  The EDCCII data presented in this table has not been
corrected for spurious detections.
\label{density}
}
\end{center}
\end{table*}

In the section, we present the results of applying the match filter
algorithm as outlined in Section \ref{method} to the whole EDSGC (as
defined in Section \ref{EDSGC}).  However, we must first establish a
common framework within which to compare our results with previous
studies. To this end, we will quote our results both as a function of
$\Lambda_{cl}$, the cluster richness as defined by Postman et al.
(1996) for the PDCS catalogue, and $R_m$, which is defined by us.  As
$R_m$ and $\Lambda_{cl}$ are just different richness normalizations of
the match filter model (see Section \ref{matched}), it is easy to
relate the two analytically.  This was achieved by integrating a
normalized Schtecter function of richness $R_m$ over the luminosity
range discussed in Section \ref{matched} and dividing by $L^{\star}$,
the characteristic luminosity of the Schtecter function. This is
equivalent to the original definition of $\Lambda_{cl}$ in Postman et
al. (1996).  Using $\alpha=-1.1$ (Postman et al. 1996), we show that
$\Lambda_{cl}=20.3$ is $R_m=100$, $\Lambda_{cl}=40.6$ is $R_m=200$ and
$\Lambda_{cl}=81.2$ is $R_m=400$.  We quote here $\Lambda_{cl}$ in the
PDCS $V_4$ passband as this is closest to the EDSGC $b_j$ passband.

In addition to comparing with the PDCS, we wish to compare with the
original Abell catalogue.  We achieve this by combining the equality
$N_R(1.0)/N_R(1.5)\simeq0.7$ (taken from Postman et al. 1996) with
Figure 17 of Postman et al. (1996) to obtain an empirical relationship
of $R_{abell} \sim 2\times \Lambda_{cl}$ for low redshift clusters
($z\le0.2$). Therefore, the three $\Lambda_{cl}$ richnesses given
above approximately correspond to Abell Richness Classes (RC) 0, 1 and
2 respectively.  These conversions are in good agreement with those
presented in Lubin \& Postman (1996 and references therein) and used
by others (Holden, private communications; Postman, private
communications). For the rest of this paper, we concentrate on the
$R_m\ge100$ systems ($\Lambda_{cl}\ge20$ or ${\rm RC}\ge0$).

In Table \ref{density}, we present the cumulative surface densities of
EDCCII clusters using the methodology outlined in this paper.  To
compute these surface densities, as a function of richness, we have
summed the number of unique clusters with richnesses of $\ge R_m$,
divided by the effective areas in Table \ref{area}.  The upper error
bars quoted in Table \ref{density} were computed by assuming all
clusters detected at a particular richness are valid cluster
candidates {\it i.e.} if we count all clusters at that richness
regardless of duplicate entries (see Section \ref{EDSGC}).  The lower
error bars were computed using clusters only detected at that richness
{\it i.e.}  they were not detected at any other $R_m$ value. We note
that these error bars are conservative and should be viewed as
boundary values.

In Table \ref{space}, we present the EDCCII space densities along with
cluster space density measurements from the PDCS, Holden et al. (1999)
and the Abell catalogue.  For the EDCCII, we computed the space
densities of clusters, as a function of richness($n (R_m)$) using the
formula

\begin{equation}
n(R_m) = \sum^{Z_{slices}} \frac{N(z_{est}, R_m)}{\int_{z_2}^{z_1}\,V(z)\times \Omega(R_m,z_{est})}, 
\label{equ1}
\end{equation}

\noindent which is a sum over all appropriate redshift slices
($Z_{slice}$).  Here, $V(z)$ is the differential cosmological volume
(per ${\rm deg^2}$), $\Omega(R_m,z_{est})$ is the effective area as a
function of redshift in Table \ref{area}, $N(z_{est}, R_m)$ is the
number of clusters detected within a redshift slice for a given
richness $R_m$, $z_1$ and $z_2$ are the limits of integration for the
redshift slice.  For $R_m\le100$, we have summed out to
$z_{est}=0.12$, while for $R_m>100$, we sum to $z_{est}=0.15$.  The
error bars on these measurements were estimated using the same
methodology as discussed above for the EDCCII cluster surface
densities.

\section{Discussion}

Recently, Holden et al. (1999) published $84$ redshift measurements
towards $16$ low redshift PDCS clusters.  From these data, Holden et
al. (1999) showed that the matched filter redshift estimate for
$z_{est}<0.5$ PDCS clusters had an error of only $\delta z\simeq0.07$;
much smaller than previously quoted by Postman et al. (1999).
Therefore, in Table \ref{space}, we present our estimates of the PDCS
space density of low redshift clusters ($0.2<z_{est}<0.6$) using the
PDCS $z_{est}$ measurements and the data given in Table 4 of Postman
et al. (1996).  We have excluded all PDCS clusters with $\sigma<3$ and
a radius of $>200$, in the ${\rm V_4}$ data, to minimize the effects
of spurious detections (see Postman et al. 1996). Therefore, these
space densities may be lower than expected since we have potentially
excluded some real clusters as well.  This approach is valid as the
true error on $z_{est}$ is now significantly smaller than our redshift
slice {\it i.e.} a substantial number of clusters will not be
scattered in, or out, of our sample because of the error in $z_{est}$.

\begin{table*}[pt]
\begin{center}
\begin{tabular}{l|cccc}
 &\multicolumn{4}{c}{Cluster Space Densities ($10^{-6}\,h^{-3}{\rm Mpc^{-3}}$)} \\ 
$n(R_m)$	& EDCCII	& PDCS	& Holden et al. & Abell	\\
\hline \hline
$100\le R_m<200$ ($\simeq 20\le \Lambda_{cl}<40$; RC$\sim$0)	&$83.5^{+193.2}_{-36.9}$ & $6.9^{+6.1}_{-3.6}$	&
          &  $11.3$  \\
$200\le R_m<400$ ($\simeq 40\le \Lambda_{cl}<80$; RC$\sim$1)	&$10.1^{+11.3}_{-4.3}$   & $12.8^{+7.6}_{-5.0}$ & $31.3^{+30.5}_{-17.1}$ &  $4.04$  \\
$R_m\ge 400$ ($\simeq\Lambda_{cl}>80$; ${\rm RC}\ge2$)	        &$2.3^{+2.5}_{-2.3}$     & $9.4^{+6.8}_{-4.7}$  & $10.4^{+23.4}_{-8.4}$  &  $1.58$  \\
\end{tabular}
\caption{Comparison of space density measurements of the EDCCII, PDCS
and Abell catalogues.  The EDCCII space densities were calculated
using Equation 1. The error bars for the EDCCII catalogue are
discussed in the text, while we quote 68\% error bars for the Holden
et al. space densities as taken from their paper.
\label{space}
}
\end{center}
\end{table*}

For the $40\le\Lambda_{cl}<80$ (RC$\sim$1) clusters, we find good
agreement between the PDCS and EDCCII space densities.  The same can
not be said for the RC$\sim$0 systems where we differ by over an order
of magnitude. We believe this discrepancy is the combination of two
effects.  First, the PDCS contains very few $\Lambda_{cl}<30$ clusters
indicating that the catalogue is possibly incomplete at these low
richnesses (the PDCS only contains 7 such systems at these low
richnesses). Secondly, the EDCCII may slightly overestimate the space
density of low richness clusters as the matched filter tends to
de--blend rich, nearby systems ($z<0.05$) into several lower richness
clusters.  Therefore, we will not discuss this inconsistency any
further but note that this measurement will be important for the next
generation of large--area CCD cluster surveys since they will possess
the volume to probe the lower richness systems at high redshift. The
EDCCII data will provide an important zero--point for such surveys
(DeepRange, Zaritsky et al. 1997).  For ${\rm RC}\ge2$ systems, the
EDCCII appears to find a factor of $\sim4$ fewer clusters than the
PDCS, however, the significance of this discrepancy is small given the
(Poisson) errors on all measurements.

In addition to checking the matched filter redshift estimates, Holden
et al. (1999) also computed the space density of PDCS clusters using a
new, and completely independent, survey selection function than that
used by Postman et al. (1996) and in this paper. Their results are
presented in Table \ref{space} and, within the errors, are in good
agreement with both our PDCS and EDCCII space density measurements.
Moreover, Holden et al. (private communication) have recently
re-calculated their measurements of the PDCS space density, as given
in Holden et al. (1999) and Table \ref{space}, but now based on a much
larger sample of galaxy redshifts (over 700 redshifts in total). They
now find $\sim 18\times 10^{-6}\,h^{-3}\,{\rm Mpc^{-3}}$ for RC=1
systems and $\sim 7\times 10^{-6}\,h^{-3}\,{\rm Mpc^{-3}}$ for ${\rm
RC}\ge2$ systems which is closer to the EDCCII space densities
presented in this paper.

In summary, we find good agreement between all three of these surveys
(PDCS, Holden et al. and the EDCCII) which combined span a redshift
range of $0.05\le z\le 0.6$. At worst, the difference between the
three surveys is $4^{+10}_{-4}$ (Table 3). This therefore justifies
our original desire to run the matched filter algorithm on the low
redshift EDSGC data since we are now comparing clusters selected in a
similar way over this entire redshift range. We have thus removed one
of the main uncertainties associated with the PDCS as we do not see a
significant difference between the low and high redshift cluster
populations (as originally highlighted by Postman et al. 1996).
Moreover, this agreement implies that there is little evolution in the
space density of optical clusters out to $z\simeq0.5$, in agreement
with results from X--ray surveys of clusters (see Nichol et al. 1997,
1999; Burke et al. 1997; Rosati et al. 1998; Vikhlinin et al. 1998;
Ebeling et al. 1997, 1999). However, we should not overstate this
claim, since the error bars on all measurements are large. In the
future, we will need large samples of clusters that span a large range
in redshift; this should be possible with the next generation of
cluster catalogues constructed from surveys like DPOSS (Gal et
al. 1999) and the SDSS (Gunn et al. 1998). We also urge the community
to adopt one cluster--finding algorithm so it can be applied to
different catalogues (at high and low redshift) consistently.

In Table \ref{space}, we present the space density of Abell clusters
taken from Postman et al. (1996 and references therein). The EDCCII
space density measurements appear to be systematically higher than the
Abell catalogue {\it e.g.}  we find $\sim7$ times as many RC$\sim$0
clusters as Abell.  However, for ${\rm RC>0}$, this discrepancy is
much less while the errors on these measurements are large. Therefore,
we must be wary about over--interpretating any claimed discrepancy
with Abell and simply note that overall, the EDCCII has lessened the
discrepancy previously claimed to be between the high and low redshift
cluster populations.

Our potential disagreement with the Abell catalogue could be due to
two factors.  First, like the EDCCII catalogue, the effective area of
the Abell catalogue could be smaller than expected (see Section
3). Secondly, the EDCCII could be finding more clusters than the Abell
catalogue at a given richness.  The first of these two factors is hard
to quantify given the subjective nature of the Abell catalogue,
however, the second factor can be addressed by cross--correlating
individual clusters in both the EDCCII and Abell catalogues. We
discuss the latter below.

In total, we detect 182 of the 324 Abell clusters in the EDCCII area,
or 56\% of them (using a matching radius of $7.5$ arcmins). This is in
good agreement with Lumsden et al. (1992) who find $\sim70\%$
match--up between their original EDCC clusters and the Abell
catalogue.  In both cases, the percentage of match--ups is independent
of richness {\it i.e.}  neither the EDCC or EDCCII appear to have
missed Abell clusters of a particular richness class (see Figure 8 of
Lumsden et al.). In addition to comparing richnesses, we have also
compared the distance estimates of our matched, and unmatched, Abell
clusters and find little correlation. Therefore, the missing
$\sim40\%$ of Abell clusters in the EDCCII catalogue appear to be
spread evenly over all Richness and Distance Classes. Finally, for
clusters in common between the EDCCII and Abell catalogues, we find no
correlation between the two difference richness estimates {\it i.e.}
$R_m$ and Abell richness or RC. This agrees with Lumsden et al. (1992)
and Postman et al. (1996) both of whom detect a large scatter between
their richness estimates and the Abell richness estimates.

In contrast, there are a total of 2109 EDCCII clusters detected in the
area given in Section \ref{EDSGC}; a factor of $\sim8$ more than
detected in the Abell catalogue over the same area (we have excluded
the supplementary Abell catalogue here and in the above analysis).
This discrepancy is lessened when we consider only $R_m\ge100$ systems
where we find 227 EDCCII clusters (however, we note that the EDCCII
only probes to $z_{est}\le0.15$, while the Abell catalogue contains
rich systems out to $z\sim0.4$). These raw numbers reflect the
differences in the space densities measurements outlined in Table
\ref{space} and highlight that a vast majority of the new EDCCII
systems are of lower richness (in the EDCCII catalogue). Although, we
do note that $143$ of the $227$ $R_m\ge100$ systems (63\%) are not in
the Abell catalogue (which corresponds to ${\rm RC\ge0}$
systems). Again, these findings agree with the original EDCC cluster
catalogue where almost 70\% of EDCC clusters were new compared to the
Abell catalogue. These two surveys therefore lend credence to the idea
(already stated by Abell 1958 and Abell et al. 1989) that the Abell
catalogue should not be used for statistical studies.

\section{Acknowledgements}

The authors would like to thank Brad Holden, Lori Lubin,
Marc Postman and Kath Romer for their assistance during the course of
this work. We also thank Chris Collins, Stuart Lumsden, Luigi Guzzo
and Harvey MacGillivray for allowing us free access to the EDSGC data.
This project was funded through NASA grant NAG5-3202 (RCN) and two CMU
Summer Undergraduate Research Grants (DAB).

\section{References}
\noindent Abell, G. O., 1958, ApJS, 3, 211\\
Abell, G. O., Corwin, H. G. Jr., Olowin, R. P., 1989, ApJS, 70, 1\\
Bahcall, N. A., Soneira, R. M., 1983, ApJ, 270, 20\\
Bahcall, N. A., Fan, X., Cen, R., 1997, ApJ, 485L, 53\\
Bower, R. G., Castander, F. J., Ellis, R. S., Couch, W. J., Boehringer, H.,
1997, MNRAS, 291, 353\\
Briel, U. G., Henry, J. P., 1993, A\&A, 271, 413\\
Burke, D. J., Collins, C. A., Sharples, R. M., Romer, A. K., Holden,
B. P., Nichol, R. C., 1997, ApJ, 488L, 83\\
%
%
Collins, C. A., Nichol, R. C., Lumsden, S. L., 1992, MNRAS, 254, 295\\
Collins, C. A., Guzzo, L., Nichol, R. C., Lumsden, S. L., 1995, MNRAS, 274, 1071\\
Collins, C. A., Nichol, R. C., Lumsden, S. L., 2000, ApJS, in preparation\\
Couch, W. J., Ellis, R. S., Maclaren, I., Malin, D. F., 1991, MNRAS,
249, 606\\
Dalton, G. B., Efstathiou, G., Maddox, S. J., Sutherland, W. J., 1992,
ApJ, 390L, 1\\
Dalton, G. B., Efstathiou, G., Maddox, S. J., Sutherland, W. J., 1994,
MNRAS, 269, 151\\
de Grandi, S., Boehringer, H., Guzzo, L., Molendi, S., Chincarini, G.,
Collins, C., Cruddace, R., Neumann, D., Schindler, S., Schuecker, P.,
Voges, W., 1999, ApJ, 514, 148\\
Dodd, R. J., Macgillivray, H. T., 1986, AJ, 92, 706\\
Ebeling, H., Edge, A. C., Fabian, A. C., Allen, S. W., Crawford,
C. S., Boehringer, H., 1997, ApJ, 479L, 101\\
Ebeling, H., Jones, L.R., Perlman, E., Scharf, C., Horner, D., Wegner, G.,
Malkan, G., Fairley, B., Mullis, C.R., ApJ, submitted (astro-ph/9905321)\\
Gal, R. R., DeCarvalho, R. R., Odewahn, S. C., Djorgovski, S. G., 
Margoniner, V. E., 1999, AJ, astro-ph/9906480\\
Gioia, I. M., Henry, J. P., Maccacaro, T., Morris, S. L., Stocke,
J. T., Wolter, A., 1990, ApJ, 356L, 35\\
Gunn, J. E., Hoessel, J. G., Oke, J. B., 1986, ApJ, 306, 30\\
Gunn, J. E., et al., 1998, AJ, 116, 3040 \\
Guzzo, L., Collins, C. A., Nichol, R. C., Lumsden, S. L., 1992, ApJ,
393L, 5\\
Heydon-Dumbleton, N. H., Collins, C. A., Macgillivray, H. T., 1989,
    MNRAS, 238, 379\\
Holden, B. P., Romer, A. K., Nichol, R. C., Ulmer, M. P., 1997, AJ, 114, 1701\\
Holden, B. P., Nichol, R. C., Romer, A. K., Metevier, A., Postman, M.,
Ulmer, M. P., Lubin, L. M., 1999, AJ, astro-ph/9907429 \\
Jones, L. R., Scharf, C., Ebeling, H., Perlman, E., Wegner, G.,
Malkan, M., Horner, D., 1998, ApJ, 495, 100\\
Kawasaki, W., Shimasaku, K., Doi, M., Okamura, S., 1998, A\&AS, 130,
567\\
Kepner, J., Fan, X., Bahcall, N., Gunn, J., Lupton, R., Xu, G., 1999,
ApJ, 517, 78\\
Kodama, T., Bell, E. F., Bower, R. G., 1999, MNRAS, 302, 152\\
Kowalski, M. P., Cruddace, R. G., Wood, K. S., Ulmer, M. P., 1984,
ApJS, 56, 403\\
Lidman, C. E., Peterson, B. A., 1996, AJ, 112, 2454\\
Loveday, J., Peterson, B. A., Efstathiou, G., Maddox, S. J., 1992, ApJ, 390, 338\\
Lubin, L. M., Postman, M., 1996, AJ, 111, 1795\\
Lumsden, S. L., Nichol, R. C., Collins, C. A., Guzzo, L., 1992, MNRAS,
258, 1\\
Lumsden, S. L., Collins, C. A., Nichol, R. C., Eke, V. R., Guzzo, L.,
1997, MNRAS, 290, 119\\
Martin, D. R., Nichol, R. C., Collins, C. A., Lumsden, S. A., Guzzo,
L., 1995, MNRAS, 274, 623\\
Nichol, R. C., Collins, C. A., Guzzo, L., Lumsden, S. L., 1992,
MNRAS.255, 21P\\
Nichol, R. C., 1993, PhD. Thesis, Univ. of Edinburgh\\
Nichol, R. C., Collins, C. A., 1993, MNRAS, 265, 867\\
Nichol, R. C., Holden, B. P., Romer, A. K., Ulmer, M. P., Burke,
D. J., Collins, C. A., 1997, ApJ, 481, 644.\\
Nichol, R. C., Romer, A. K., Holden, B. P., Ulmer, M. P., Pildis,
R. A., Adami, C., Merrelli, A. J., Burke, D. J., Collins, C. A., 1999,
ApJ, 521, 21\\
Olsen, L. F., Scodeggio, M., Da Costa, L., Benoist, C., Bertin, E.,
        Deul, E., Erben, T., Guarnieri, M. D., Hook, R., Nonino, M.,
        Proni, I., Slijkhuis, R., Wicenec, A., Wichmann, R., 1999,
        A\&A, 345, 681\\ 
Ostrander, E. J., Nichol, R. C., Ratnatunga, K. U., Griffiths, R. E.,
        1998, AJ, 116, 2644\\
Postman, M., Huchra, J. P., Geller, M. J., 1992, ApJ, 384, 404\\
Postman, M., Lubin, L. M., Gunn, J. E., Oke, J. B., Hoessel, J. G.,
Schneider, D. P., Christensen, J. A., 1996, AJ, 111, 615 (PDCS)\\
Postman, M., Lauer, T. R., Szapudi, I., Oegerle, W., 1998, ApJ, 506, 33\\
M. Ramella, M., Nonino, M., Boschin, W., Fadda, D., 1998, astro-ph/9810124\\
Reichart, D. E., Nichol, R. C., Castander, F. J., Burke, D. J., Romer,
A. K., Holden, B. P., Collins, C. A., Ulmer, M. P., 1999, ApJ, 518,
521\\
Romer, A. K., Nichol, R. C., Holden, B. P., Ulmer, M. P., Pildis,
R. A., Adami, C., Merrelli, A. J., Kron, R., Burke, D. J., Collins,
C. A., 1999, ApJS, accepted\\
Rosati, P., Della Ceca, R., Norman, C., Giacconi, R., 1998, ApJ, 492L,
21\\
Schuecker, P., Boehringer, H., 1998,A\&A, 339, 315\\
Slezak, E., Bijaoui, A., Mars, G., 1990, A\&A, 227, 301\\
Vikhlinin, A., McNamara, B. R., Forman, W., Jones, C., Quintana, H.,
        Hornstrup, A., 1998, ApJ, 502, 558\\
Zaritsky, D., Nelson, A. E., Dalcanton, J. J., Gonzalez, A. H., 1997,
        ApJ, 480L, 91\\

\end{document}